\documentclass[11pt, a4paper]{article}

\usepackage{jcappub}

\usepackage{graphicx}
\usepackage{datetime}
\usepackage{textpos}
\usepackage{booktabs}

\title{Gamma-ray Lines in the Fermi Data: is it a Bubble?}
\author{Stefano Profumo$^{1,2}$ and Tim Linden$^{1}$}
\affiliation{$^1$ Department of Physics, University of California, Santa Cruz, 1156 High Street, Santa Cruz, CA, 95064}
\affiliation{$^2$ Santa Cruz Institute for Particle Physics, University of California, Santa Cruz, 1156 High Street, Santa Cruz, CA, 95064}
\emailAdd{profumo@ucsc.edu}
\emailAdd{tlinden@ucsc.edu}

\abstract{Recently, tentative evidence for an excess of gamma rays at energies around 130 GeV has been reported from analyses of data from the Fermi Large Area Telescope (LAT). The excess is potentially of great interest, as it could be associated with the pair-annihilation of Galactic dark matter and  the subsequent production of monochromatic or internal bremsstrahlung gamma rays. The 130 GeV excess appears when an optimized selection of the target region of interest is employed, a procedure that depends upon the assumed dark matter density profile. For the profiles producing an appreciable signal, these target regions vastly overlap with the region corresponding to the so-called ``Fermi bubbles''. We argue that the tentative evidence for a line feature is likely due to hard photons in the Fermi bubbles regions, where the gamma-ray spectrum contains a spectral break in the energy range of interest (100 -- 150 GeV). Although the origin of this broken power-law is unclear, it is probably related to standard astrophysical processes and not to dark matter annihilation.  A broken power-law provides as good a fit as a line ``excess'', even within small energy windows, and a significantly better fit for large energy windows.}

\begin{document}
\maketitle

\section{Introduction}
\label{sec:introduction}
Gamma rays provide a prime search channel for a signature from the pair-annihilation or the decay of Galactic dark matter. Data from the first two years of operation of the Fermi gamma-ray Large Area Telescope (LAT), however, have shown that no clean and unambiguous signal from dark matter is present in the gamma-ray sky. The most recent constraints from LAT observations of dark-matter dominated Milky Way satellite dwarf spheroidal galaxies have started to probe cross sections that, in particle models, lead to thermal production of dark matter in the early universe \citep{dsph}. The central problem in ascertaining that a gamma-ray signal originates from dark matter annihilation is the absence of clear spectral features in the predicted dark matter pair-annihilation spectrum: if dark matter annihilates primarily into hadronic final states, the resulting gamma rays are produced by a featureless neutral pion two-photon decay process. Similarly, annihilation into leptons leads to featureless power-law spectra, with possible edges at gamma-ray energies corresponding to the dark matter particle mass -- where statistics is usually very poor.

An exception to this rule applies to models where the dark matter pair-annihilates into one or two (monochromatic) gamma rays directly, or to processes leading to the final state radiation of hard photons\footnote{The latter case includes exotic instances such as final state radiation from dark matter-cosmic-ray scattering, \cite[see e.g.][]{ubaldi,tait,Profumo:2011jt}.}. Generically, since the dark matter is not electromagnetically charged, these pair-annihilation processes are loop-suppressed, and, typically, (for example for neutralino dark matter in supersymmetry) they have branching ratios on the order of $10^4$ times smaller than those leading to other, unsuppressed final states \citep[see e.g. ][]{ullio}. On the up-side, a gamma-ray line or a sharp edge are very distinct spectral features. On the down-side, given the absence of any convincing signal of dark matter annihilation in the gamma-ray continuum, these distinctive spectral features are also expected to be extremely faint.

The Fermi-LAT Collaboration searched for a gamma-ray line signal in 11 months of data, producing a null result \citep{fermiline}. The targeted region of interest (ROI) was not, however, optimized to produce the largest possible signal-to-noise: rather, it covered most of the sky, albeit removing the galactic plane ($|b|>10^\circ$), while including a 20$^\circ\times20^\circ$ square centered at the Galactic center. An updated analysis, also from the Fermi-LAT Collaboration, based on 24 months of data, and employing the same ROI but tighter photon selection cuts, also resulted in a null result \cite{newline}, producing, as expected, even more stringent limits than the previous study. More recently, however, two studies have carried out a more refined, but model-dependent determination of the optimal ROI for given dark matter density profiles (to be described below), and searched for the sharp peak expected from virtual internal bremsstrahlung (IBS) in certain dark matter models \citep{ibs}, and for a monochromatic gamma-ray line \citep[hereafter W12]{line}. These two studies employed significantly larger photon  statistics (corresponding to 43 months of data) than the Fermi-LAT Collaboration study, but employed the same spectral analysis: a sliding energy window (of varying size) and a power-law fit to the ``background'' within the window. Interestingly, tentative evidences for spectral features corresponding to an IBS signal, possibly associated to a dark matter mass of about 150 GeV (with a significance of 4.3$\sigma$, reduced to 3.1$\sigma$ including the look-elsewhere effect) have been reported in \cite{ibs}; an even stronger hint, this time for a gamma-ray line signal at about 130 GeV (with a significance of 4.6$\sigma$, reduced to 3.3$\sigma$ including the look-elsewhere effect) was reported in W12.

\begin{figure}
                \centering
  \includegraphics[width=.6\linewidth]{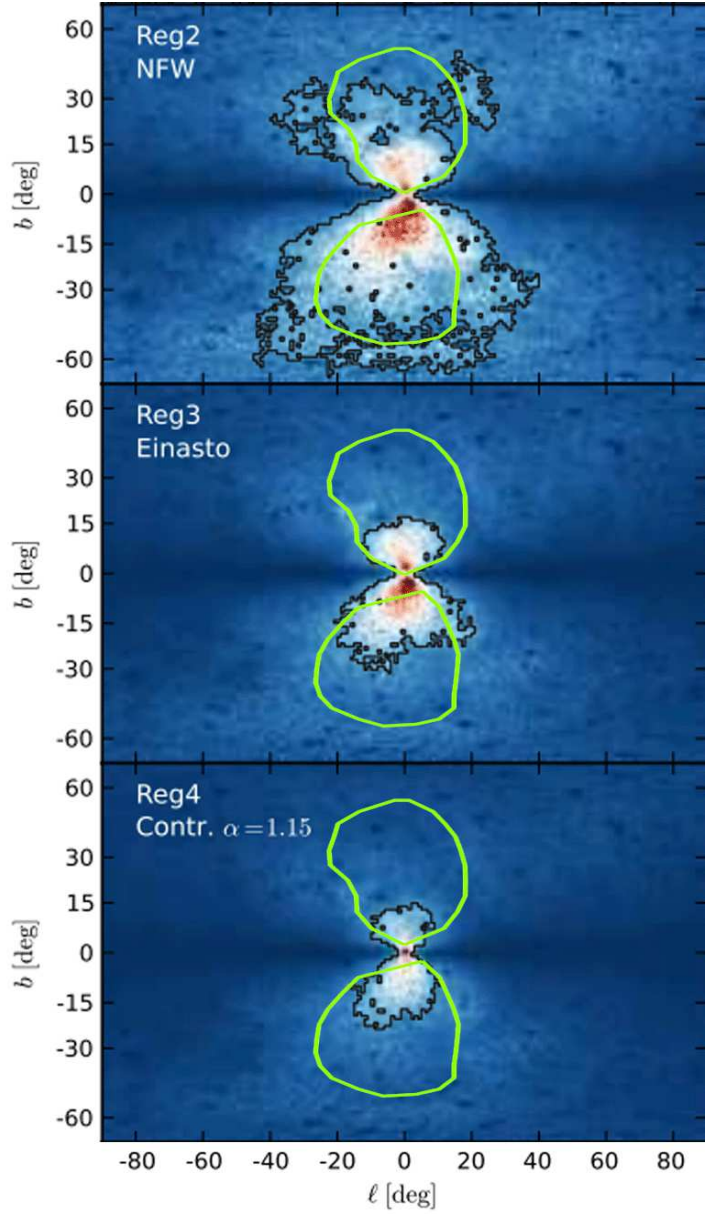}
                	\caption{ \label{fig:overlay} The target ROI's employed in \cite{line} (reproduced here with permission of the Author), overlaid with the contours corresponding to the Fermi bubbles regions as determined in \cite{fermibubbles}. The three panels refer to three different assumptions for the dark matter density profile, specifically an NFW Profile (top), an Einasto profile (middle) and an NFW with an inner power-law slope r$^{-1.15}$ (bottom).}
\end{figure}
The key novelty of the two recent analyses \citep[W12]{ibs} is not in the spectral analysis, but in the optimization of the ROI. For a given Galactic dark matter density profile, the target region is chosen to maximize the signal-to-noise; for a given direction in the sky, the signal is defined as the integral over a given line of sight of the density profile squared, while the noise is defined as the square-root of the number of photons in the 1-20 GeV energy range in the given direction, within a certain angular region. As a result, cored profiles tend to produce ROI's extending over very large angular regions, while profiles with a steeply rising central density select preferentially regions closer to the Galactic center, excluding the Galactic plane where the ``noise'' is especially large. We reproduce in Fig.~\ref{fig:overlay} the resulting regions (taken from W12, with permission of the Author) for three dark matter density profiles (respectively a Navarro-Frenk and White (NFW) profile for Reg2, an Einasto profile for Reg3 and a generalized NFW profile, with an inner density proportional to $r^{-1.15}$, with $r$ the radial distance from the Galactic center, for Reg4 -- see W12 for details). The three regions correspond to those where tentative evidence for a gamma-ray line was found in W12.\\[0.5cm]

\section{A Gamma-Ray Line and the Fermi Bubbles}
As a result of the procedure determining the ROI in \cite{ibs} and in W12, the selected regions in the sky overlap very significantly with the region where giant gamma-ray bubbles (known as the ``Fermi bubbles'') have been recently discovered \citep{fermibubbles}. In order to illustrate this point we overlay in Fig.~\ref{fig:overlay}  the bubbles' contours, as determined by \cite{fermibubbles}, directly on the three ROIs determined in W12 that lead to the most significant line excess. It is clear that most of the highest signal-to-noise regions used by W12 are contained within the Fermi bubbles. As we discuss below, this is a crucial observation, since the Fermi bubbles have been found to host a hard gamma-ray spectrum, with a possible spectral break at energies between 100 and 200 GeV. This spectrum is significantly distinct from the (softer) power-law spectrum expected from inverse Compton or hadronic processes \citep{fermibubbles}, and from the power-law observed in the isotropic gamma-ray component in the Fermi-LAT sky \citep{isotropic}. The gamma-ray emission associated with the bubbles is in general a complicating factor in using diffuse gamma-ray data as a probe of dark matter \citep[as explicitly pointed out already in][]{fermibubbles}. Here, we show that indeed the hard emission from the Fermi bubbles can be responsible for, or at the least is a very significant contaminant to, the claimed evidence for a gamma-ray line reported in W12.

The key features of the Fermi bubbles are (i) a very broad extension in the sky, about 50$^\circ$ above and below the Galactic center with a maximum width of 40$^\circ$, (ii) the previously mentioned hard spectrum ($dN/dE\sim E^{-2}$), and (iii) the absence of spatial variation in the gamma-ray spectrum throughout the bubble regions or between the North and South bubbles \citep{fermibubbles}. In addition, sharp edges are observed corresponding to the contours reproduced in Fig.~\ref{fig:overlay}. Although several hypotheses have been put forward to explain the origin, spectrum and morphology of the Fermi bubbles, no consensus has been reached so far. Astrophysical processes that might be related to the production of the hard excess gamma-ray emission observed in the Fermi bubbles region include recent AGN jet activity in the Galactic center \citep[e.g.][]{guo}, star capture by the central super-massive black hole \citep[e.g.][]{2011ApJ...731L..17C}, and a population of high-energy confined cosmic-ray protons \citep[e.g.][]{Crocker:2010dg, Crocker:2010qn}. As argued in the original work of \cite{fermibubbles}, a dark matter interpretation for the bubbles is unlikely, especially based on spectral features, morphology and the associated radio and X-ray counterparts. Contrived dark matter density profiles and/or diffusion models must be invoked to explain the bubbles' gamma-ray spectrum with dark matter annihilation \citep[see e.g.][]{2011ApJ...741...25D}. Finally, we note that the gamma-ray flux extracted by \cite{fermibubbles} (see e.g. their Figure 14) shows an extremely hard spectrum, which softens sharply at energies of around 110-150~GeV. 

\begin{figure}
                \centering
  \includegraphics[width=\linewidth]{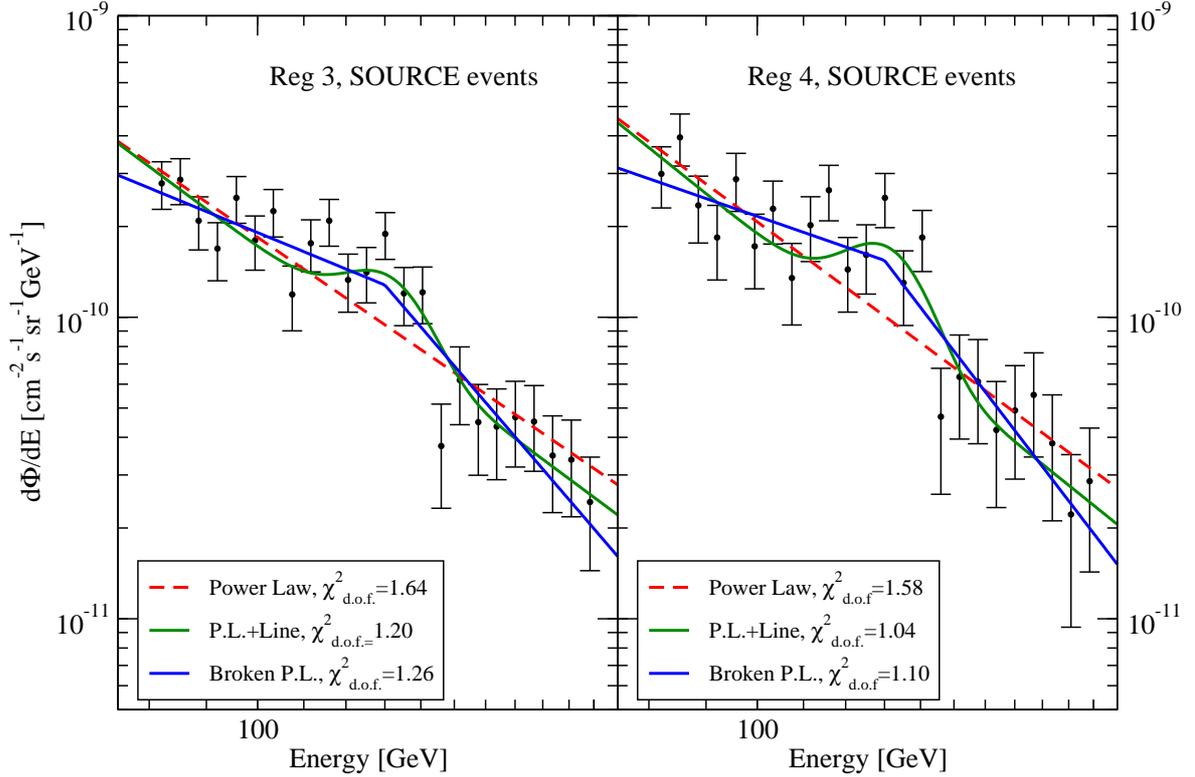}
                	\caption{ \label{fig:spectrum}The gamma-ray spectrum in Regions 3 (Einasto Profile, left) and 4 (r$^{-1.15}$ profile, right), from the analysis of \cite{line}, and the best-fit power-law (red dashed), power-law plus monochromatic line at $E_\gamma=130$ GeV, smeared by the LAT energy resolution (green), and a broken power-law (blue) similar to the spectrum observed in the Fermi bubbles regions.}
\end{figure}

The sudden softening of the Fermi bubbles spectrum at an energy of approximately 150~GeV~\cite[Fig. 14]{fermibubbles} is important, as this spectral break can easily be confused with a gamma-ray line. In general, any attempt to match a broken power-law spectrum by a single power-law produces a best fit spectrum with a spectral index intermediate between the two spectral indices from the best-fit broken power-law. In the case which is relevant here, where the high energy spectral index is softer than the low energy index, this creates a deficit in medium energy emission which is always most pronounced at the point of the spectral break. Neglecting complicating factors such as an energy dependence in the statistical error of the input function, the addition of a delta-function ``bump" will provide the largest improvement to the overall fit when the delta-function resides at the point of the spectral break. Adding a ``line'' will produce an increasingly more beneficial effect to the quality of the fit the narrower the energy window under consideration. Notice that the TS method employed in W12 in fact picks up several apparent spectral breaks, as can be seen by comparing the spectra shown in the right panels of Figure 1 of W12 with the results shown in Figure 2 of W12 at energies around 25, 40, 50 and, naturally, with much higher significance, 130 GeV.

We start by considering the fluxes presented in App.~A of W12, restricting our analysis to the sliding energy window where the most highly statistically significant line is claimed (i.e. approximately between 80 and 200 GeV). We find that a spectrum similar to that inferred for the hard emission in the Fermi bubbles provides a goodness of fit to the gamma-ray data which is of comparable quality to that obtained by adding a gamma-ray line. This is illustrated in Fig.~\ref{fig:spectrum}, where we show fits to the data quoted in W12 and obtained from the SOURCE events selections (Pass 7 Version 6) for Regions 3 and 4. We employ a power law (red dashed line), a power law plus a line at 130 GeV (smeared by the energy resolution of the LAT at 130 GeV, solid green line), and a broken power-law (break set at 130 GeV, solid blue line). We obtain slightly better $\chi^2$ per degree of freedom in the case of the power-law plus line with respect to the broken power-law (1.20 versus 1.26 for Reg3, and 1.04 versus 1.10 for Reg4). However, this difference is not statistically significant. Note that although we let the broken power-law slopes vary freely, we obtain spectra that are compatible with those found in a similar energy range by \cite{fermibubbles} for the Fermi bubbles. This provides a preliminary indication that the gamma rays identified as originating from a monochromatic 130 GeV emission might potentially stem from the astrophysical process that produces the hard spectrum observed in the Fermi bubbles region.

An important aspect of the spectral fitting procedure is the size of the energy window. While this should exceed the instrumental energy resolution, the signal should be resilient to changes in the size of the ``sliding'' energy window. Indeed, W12 demonstrates this point in a compelling manner with Figure 8, where it is shown that the TS value is essentially stable as a function of $\epsilon$, where the energy windows centered around $E_\gamma=E_0$ are $[E_0/\sqrt{\epsilon},\ {\rm min}(E_0\sqrt{\epsilon},\ 300\ {\rm GeV})]$. Here we argue that while increasing the energy window does not reduce the TS for the fit with a line, it does increase the quality of the fit with a broken power-law. This seems to indicate that when a larger portion of the spectrum is taken into consideration, the line appears more and more like the edge of a broken power-law, rather than as a monochromatic gamma-ray line. We consider this to be an additional hint pointing towards a Fermi bubbles origin for the hard photon events.

\begin{figure}[!t]
                \centering
  \includegraphics[width=.6\linewidth]{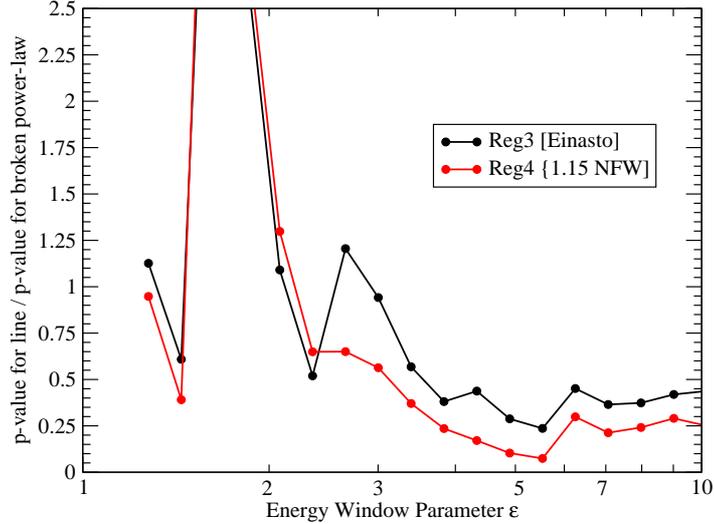}
                	\caption{ \label{fig:enewindow} The ratio of the p-values for a fit with a power-law plus a line and for a fit with a broken power-law, as a function of the energy window size (the parameter $\epsilon$ is defined, as in W12, such that the energy window is $[E_0/\sqrt{\epsilon},\ {\rm min}(E_0\sqrt{\epsilon},\ 300\ {\rm GeV})]$).}
\end{figure}

We illustrate this point with Fig.~\ref{fig:enewindow}, where we show the ratio of the p-value corresponding to a fit with a power-law plus a line, over the p-value for a broken power-law, in Regions 3 (Einasto, black line) and in Region 4 (NFW with a 1.15 slope, red line) as a function of $\epsilon$, from the data in Fig.~1, right panels, of W12. A ratio smaller than one indicates that a better fit is achieved with a broken power-law with a break at 130 GeV, while a ratio larger than one indicates that a better fit is obtained for a gamma-ray line at 130 GeV. Barring the narrow windows of the first few points, the ratio stabilizes well below 1 for $\epsilon\gtrsim2$. Enlarging the energy window clearly indicates that the hard photons likely originate from a broken power-law, whose slopes cannot be correctly reconstructed using small energy windows (therefore leading to a spuriously better fit with a gamma-ray line). We note that the details of the results shown in Fig.~\ref{fig:enewindow} might depend on binning (although we do not expect the main conclusion to be affected); an un-binned analysis might be warranted.

An additional test to the hypothesis put forward here stems from considering the spectrum of gamma rays from the portion of the W12 ROIs {\em which are outside} the Fermi bubble regions. For the ROI's Reg3 and Reg4, where the highest line significance is found, these outlying regions correspond to a low-galactic latitude strip with approximately $|b|,|l|\lesssim 5^\circ$. This highly gamma-ray bright region is not expected to exhibit any significant signal that could be associated with a bump in the 130 GeV energy range, and is, rather, likely characterized by a featureless power-law spectrum, since:

(i) no signal appears in Reg5 of W12 or in Reg4 of \cite{ibs}, which both envelope similar regions to the portions of Reg3 and Reg4 in W12 which lie outside of the Fermi bubbles; the quoted spectra there are compatible with a featureless power-law;

(ii) studies of the inner Galaxy \citep[e.g.][and references therein]{Cholis:2009gv, Vitale:2009hr} do not indicate any excess compatible with the 130 GeV line signal of W12; all of these studies indicate a featureless power-law extending to the largest energies under consideration;

(iii) the template analysis of \cite{fermibubbles} indicates that in the low-Galactic latitude region the hard photon excess characterizing the Fermi bubbles disappears, or is buried under the bright, power-law Galactic background.

The lines of evidence above indicate that a featureless power-law spectrum is expected from the ROI's portions lying outside the Fermi bubbles. To conclusively test this point, however, a new dedicated Fermi-LAT data analysis is warranted, albeit this is outside the scopes of the present note.

\begin{figure}[!t]
               \centering
  \includegraphics[width=.6\linewidth]{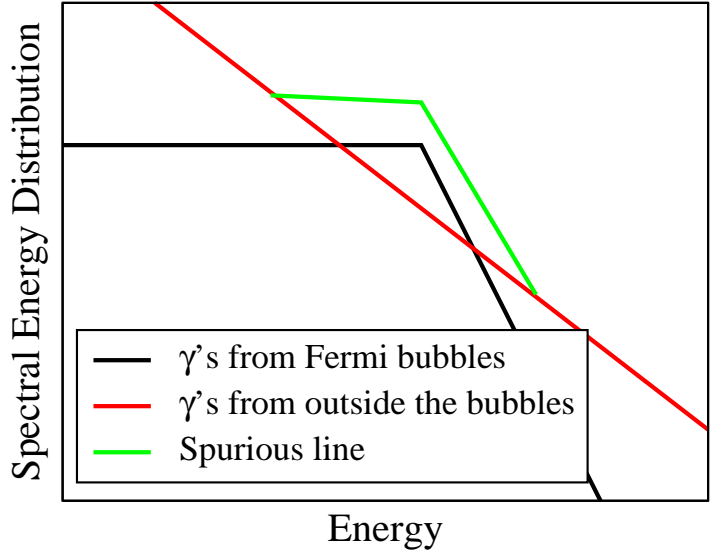}
                	\caption{ \label{fig:schematic} A schematic picture of why it is likely that a broken power-law spectrum from the Fermi bubbles region appear as a spurious line if a bright, featureless power-law background from ROI portions outside the Fermi bubbles is added.}
\end{figure}

Lastly, it is claimed in footnote 6 of W12, that utilizing the Fermi bubbles template does not produce any signal (the corresponding ROI and TS are therefore not shown in W12 and deemed to be not among ``the most interesting ones''). Is this consistent with the present interpretation, that associates the tentative signal precisely to photons from the Fermi bubbles? The answer is yes. Fitting the gamma rays from the Fermi bubble regions, a broken power-law provides clearly a much better spectral description than a line, as evident in the analysis of \cite{fermibubbles}. As we argue above, the ROI portions outside the Fermi bubble regions likely do not contain any significant signal, and have featureless power-law spectrum. We therefore expect that contaminating the Fermi bubble regions with photons from the bright central regions (for Regions 3 and 4) or with large regions dominated by Galactic and extragalactic background and point sources (Region 2), leads to ``watering down'' the Fermi bubble photons with a bright power-law background. As a result, the clear power-law break in the Fermi bubbles spectrum is flooded with background photons with a simple power-law spectrum: the bubbles spectral break will then appear as a ``tip of the iceberg'' bump, easily misinterpreted as a line. We schematically illustrate why we expect a spurious line signal from the W12 ROI's but not from the Fermi bubbles with Fig.~\ref{fig:schematic}.

\section{Discussion and Conclusions}
\label{sec:conclusions}
In this note we argue that the tentative evidence for a gamma-ray line possibly associated with dark matter pair-annihilation reported in W12 is due to excess hard photons from the Fermi bubbles regions, where the gamma-ray spectrum is a broken power law. Although the origin of this broken power-law is unclear, it probably has to do with standard astrophysical processes and not with dark matter annihilation. 

Our claim is substantiated by two key observations:

(I) The regions where the tentative gamma-ray line signal is significant largely overlap with the Fermi bubble regions (Fig.~\ref{fig:overlay}); there are several reasons why we expect a featureless power-law gamma-ray spectrum (and no evidence for a signal) from the portions of the ROI's that do not overlap with the Fermi bubble regions. Given the gamma-ray spectrum in the Fermi bubbles, a spurious line is then likely to emerge from data including the full ROI, as illustrated in Fig.~\ref{fig:schematic}.

(II) The power-law break in the Fermi bubbles occurs in the 110-150 GeV region \citep{fermibubbles}. Due to the relatively poor energy resolution at those energies, such spectral break could be confused, in a spectral analysis employing a sliding energy window, for a gamma-ray line signal. We find that a broken power-law gives as good a fit as a line, even in a small energy window (Fig.~\ref{fig:spectrum}); widening the size of the energy window points quite conclusively towards a broken power-law over a line fit (Fig.~\ref{fig:enewindow}).

As observed by others, the spectrum and morphology of the Fermi bubbles are inconsistent with general expectations for a dark matter annihilation signal, and the hard gamma rays from those regions are likely of standard astrophysical origin. From W12 and \cite{ibs} we derive the important lesson that the regions of the sky where the ``signal-to-noise'' is maximal, for several well-motivated dark matter density profiles, overlap largely with the Fermi bubbles (Fig.~\ref{fig:overlay}). We thus conclude that the Fermi bubbles are a crucial background to any search for a diffuse (or for a faint, as in the case of a line) gamma-ray signal from dark matter annihilation.

Recently, a new and independent re-analysis of the Fermi-LAT ULTRACLEAN dataset was performed in Ref.~\cite{tempetal}. This study adopts a spectrally uniform background model for the high-energy gamma-ray sky -- close to a 2.6 power-law, extrapolated by cutting out the central ${12}^\circ$ of the Galaxy -- and makes use of a kernel smoothing method to fit the resulting excess ``signal''. Ref.~\cite{tempetal} claims to find a 4.5$\sigma$ peak at 130 GeV from the central regions of the Galaxy, as well as a number of other regions with significances as high as 3.2$\sigma$, tentatively identified with ``the first DM sub-haloes of our Galaxy''. 

Ref.~\cite{tempetal} also argues against the possibility, which we suggest here, that photons from the Fermi bubbles significantly contribute to the signal reported in W12. We note that this last claim is based on the finding that the majority of photons in the $120<E_\gamma<140$ GeV range (fig.~2, left, in Ref.~\cite{tempetal}) mostly originate from the Galactic plane, thus outside the Fermi bubble regions of Ref.~\cite{fermibubbles}. What is shown in Ref.\cite{tempetal}, however, only proves that the Galactic plane hosts most of the $120<E_\gamma<140$ GeV photons -- a fact which is well known. It does not disprove that the majority of the photons contributing to the observation of a $\gamma$-ray bump in W12 (and therefore in the W12 ROI's, that cut out most of the Galactic plane) is associated with the Fermi bubbles. In addition, it is important to note that the Fermi bubbles potentially extend beyond the ``bubble regions'' of \cite{fermibubbles} into the Galactic center region. Most of the high-energy photons from the bubbles are in fact probably from the innermost regions of the Galaxy, the only region where e.g. the X-ray counterpart to the gamma-ray bubbles has been identified \cite{fermibubbles}.

While it is important to have additional, independent analyses of the claimed monochromatic excess, we have reservations about the background modeling procedure adopted in Ref.~\cite{tempetal}, and thus about the conclusions of that study. First, most of the claimed signal originates from the portion of the sky that is not used to model the background (the innermost ${12}^\circ$ Galactic center region); 
Second, no point-source subtraction was adopted in the study. This is particularly concerning for the~\cite{tempetal} study, as they examine 3$^\circ$ ROIs which show evidence for a $\gamma$-ray line. In this case, contributions from a single bright point-source (possibly with a spectrum not-well modeled by the power-law background) could create a bump-like residual. Indeed we note that the two most statistically significant excesses identified by \cite{tempetal} (the Central and West Regions) are likely to have large contributions from known Fermi-LAT point sources. Specifically, the Central Region lies $\sim$1$^\circ$ from 2FGL J1745.6-2858, which is the 9th brightest point source in the 2FGL catalog, and which has a spectral index of 2.34, significantly harder than the 2.6 background assumed in this work. The West Region lines within 0.5$^\circ$ of  2FGL J1718.1-3725, which has an extremely hard 1.99 spectral index, implying its contribution to the high energy signal may be considerable.

Finally, we note that additional improvements in the understanding of the Fermi-LAT instrumental sensitivity are expected which will greatly improve our ability to differentiate between a photon line signal as reported in Ref.~\cite{line} and the broken-power law interpretation supported here. Among the expected improvements in the upcoming Pass~8 version of the Fermi-LAT analysis software is a greatly enhanced effective area for high energy gamma-rays as well as a significantly improved energy resolution~\citep{pass8}, which are two to the most critical ingredients in the differentiation between the line and the broken power-law models described here. Future work on the regions of the gamma-ray sky investigated here and in W12 is therefore imperative.

\acknowledgments
We are grateful to Andrea Albert, Bill Atwood, Eric Charles, Johann Cohen-Tanugi, Tesla Jeltema and Christoph Weinger for helpful comments and discussions. This work is partly supported by NASA grant NNX11AQ10G. SP acknowledges support from an Outstanding Junior Investigator Award from the Department of Energy, and from DoE grant DE-FG02-04ER41286. \\ 


\providecommand{\href}[2]{#2}\begingroup\raggedright\endgroup

\end{document}